# Isotropic Scaling Features Measured Locally in the Solar Wind Turbulence with Stationary Background Field

Honghong Wu[1], Chuanyi Tu[1], Xin Wang[2], Jiansen He[1], Liping Yang[3], and Linghua Wang[1]
[1] School of Earth and Space Sciences, Peking University, Beijing, People's Republic of China; chuanyitu@pku.edu.cn
[2] School of Space and Environment, Beihang University, Beijing, People's Republic of China
[3] SIGMA Weather Group, State Key Laboratory for Space Weather, National Space Science Center, Chinese Academy of Sciences, Beijing, People's Republic of China

## Abstract

The scaling anisotropy is crucial to interpret the nonlinear interactions in solar wind turbulence. Previous observations provide diverse results and the structure function analyses are also reported to be an approach to investigate the scaling anisotropy based on a local magnetic field. However, the determination of the sampling angle with respect to the local background magnetic field requires that the observed time series for the average are time stationary. Whether or not this required time stationarity is compatible with the measurements has not been investigated. Here we utilize the second-order structure function method to study the scaling anisotropy with a time-stationary background field. We analyze 88 fast solar wind intervals each with time durations ⩾ 2 days measured by *WIND* spacecraft in the period 2005–2018. We calculate the local magnetic field as the average of the time series $\boldsymbol{B}(t')$ whose time stationarity is fulfilled by our criterion $\phi < 10°$ ($\phi$ is the angle between the two averaged magnetic fields after cutting $\boldsymbol{B}(t')$ into two halves). We find for the first time the isotropic scaling feature of the magnetic-trace structure functions with scaling indices $-0.63 \pm 0.08$ and $0.70 \pm 0.04$, respectively, with the local magnetic field parallel and perpendicular to the solar wind velocity directions. The scaling for the velocity-trace structure functions is also isotropic and the indices are $-0.47 \pm 0.10$ and $0.51 \pm 0.09$. We also find that with increasing $\phi$ threshold to 90°, the scaling index of the magnetic-trace structure function in the parallel direction decreases to $-0.81$, while the rms of the instantaneous angle between magnetic field and solar wind velocity increases up to 45° at the timescale 150 s, indicating a mix of perpendicular measurements into parallel ones at large scales.

*Unified Astronomy Thesaurus concepts:* Solar wind (1534); Interplanetary turbulence (829); Magnetic fields (994); Space plasmas (1544)

## 1. Introduction

The scaling anisotropy is one of the important issues regarding the energy cascade processes in the solar wind turbulence (Tu & Marsch 1995; Horbury et al. 2012; Bruno & Carbone 2013). The observational results on this property are somewhat diverse and not fully understood. Tessein et al. (2009) found that the power spectral index of magnetic fluctuations and velocity fluctuations is about $-5/3$ and $-1.4$ respectively. In their work, no statistically significant dependence on $\theta_{\rm RB}$ is seen from both the structure function analysis and the trace of power spectral analysis and $\theta_{\rm RB}$ is the sampling angle with respect to the global mean magnetic field. Wang et al. (2015) presented that the power spectral index of the magnetic fluctuations are $-1.67$ and $-1.46$, respectively, for small and large sampling angles also with respect to the the global mean magnetic field. They selected the time periods that satisfy the conditions of stationarity and the fluctuations are low amplitude. However, since the anisotropy is a local property, one needs to calculate the local mean magnetic field (Podesta 2009).

Horbury et al. (2008) first presented, based on wavelet analysis, the observational spectral anisotropy considering the local mean magnetic field, which is calculated as the sum over the data set of the product of the amplitude envelope of the Morlet wavelet with the magnetic field time series. They found the spectral index is $-2$ for the parallel fluctuations. Several later works used the same local mean magnetic field with different data sets and confirmed this anisotropy feature (Podesta 2009; Luo & Wu 2010; Chen et al. 2011; Wicks et al. 2011). The anisotropic pattern of 2D PSD in the MHD wavevector ($k_\parallel$, $k_\perp$) space was reconstructed for the first time by He et al. (2013) with their developed method of multi-dimensional PSD's tomography. The 2D magnetic PSD($k_\parallel$, $k_\perp$) shows an oblique power ridge bending more to $k_\perp$ as the dominant population. Such a pattern of magnetic PSD($k_\parallel$, $k_\perp$) are confirmed again by Yan et al. (2016), who also illustrated the similar anisotropic pattern of velocity PSD($k_\parallel$, $k_\perp$) indicating the evidence of prevailing oblique Alfvénic waves. Magnetic power is dominant over velocity power in the subregion near ($k_\parallel \sim 0$, $k_\perp$), manifesting the contribution from magnetic structures.

However, Wang et al. (2014) applied the same method as Horbury et al. (2008) but removed the intermittency. The spectral index of magnetic field becomes $-1.63$ and does not depend on the sampling angle, suggesting that the spectral anisotropy of solar wind power spectra in the inertial range could result from turbulence intermittency. Furthermore, the multiorder structure functions after removing the intermittency are found to be angle-independent, and show the linear trend of scaling exponent as a function of order (i.e., mono-fractal feature; Pei et al. 2016). Wang et al. (2016) performed the wavelet analysis but selected the data with a stricter criterion to better ensure the time stationarity during the local sampling interval and the spectral index of magnetic field are found to be −1.75 and −1.65 for the parallel and perpendicular fluctuations, respectively.

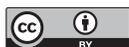

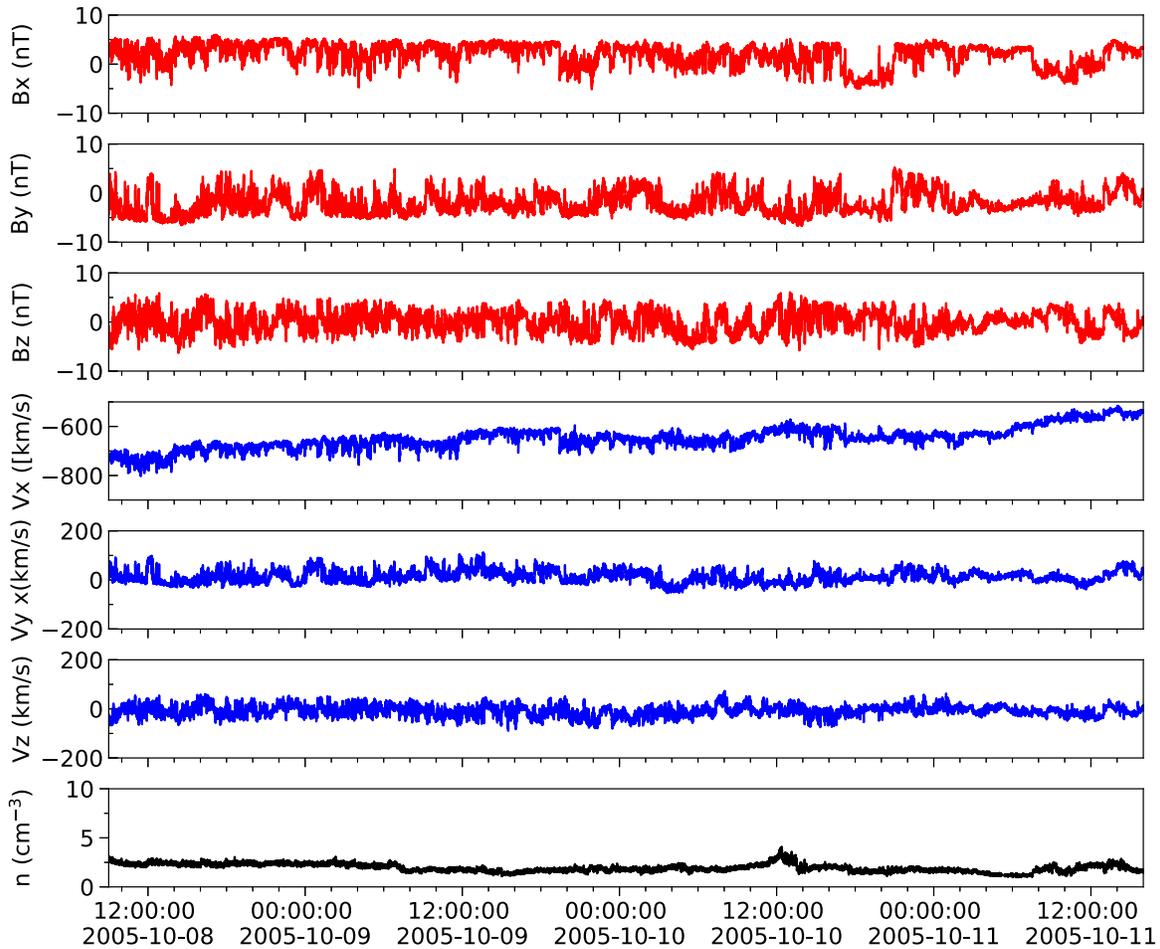

**Figure 1.** Interval in the fast solar wind (2005 October 8 09:00:00 UT−2005 October 11 16:00:00 UT) observed by *WIND* spacecraft. The red lines are magnetic field components in GSE coordinates. The blue lines are velocity components in GSE coordinates. The black line is proton number density.

Cho & Vishniac (2000) first introduced the local mean magnetic field as the average of magnetic field measured at two space points or two time instants for the structure function study. The difference between the two-point magnetic field vectors is used to calculate the second-order structure functions. The structure function index is equal to the spectral index when subtracted by 1. Later on, several works investigated the scaling of the structure function using the same method (Chen et al. 2012; Mallet et al. 2016; Verdini et al. 2018; Yang et al. 2018). These works believe that the measurements at every two points in the solar wind represent turbulence variations, but that they are not influenced by convective structures, and that the magnetic field averaged between the two-point measurements represents the major instantaneous direction during the two time instants and thus can be used to check the orientation of the measurements exactly. However, in the solar wind measurements, the convective structures and large variations may have an impact on the averaged magnetic fields, and the local mean magnetic field may not represent the major direction of the individual measurement. To avoid these influences one needs to guarantee the time stationarity of the time series between the two measurement points. However, this time stationarity has not been studied before. In this study, we perform the scaling anisotropy analysis based on the structure functions for both magnetic field and velocity using fast wind measurements from *WIND* spacecraft. The local mean magnetic field is calculated as the average of magnetic field time series measured between two time instants. We require the magnetic field series to be time stationary. In Section 2, we describe the data and methods in detail. In Section 3, we analyze the criterion of time stationarity. We show our results in Section 4 and discuss our results and present our conclusions in Section 5.

## 2. Data and Method

In the analysis, we use fast solar wind data from *WIND* spacecraft. The magnetic field data and the plasma data both with time resolution $\Delta = 3$ s are from the magnetic field investigation (Lepping et al. 1995) and the three-dimensional plasma analyzer (Lin et al. 1995), respectively. Figure 1 shows a fast solar wind interval with the velocity $|V(t)| > 500$ km s$^{-1}$ and the proton number density $n(t) < 5$ cm$^{-3}$ at every time instantaneous $t$ during 2005 October 8 09:00:00 UT−2005 October 11 16:00:00 UT. We select all the fast solar wind intervals each with time durations ⩾ 2 days from 2005 to 2018 when *WIND* spacecraft hovers at the Lagrangian point L1 using the criteria that the solar wind velocity $|V(t)| > 500$ km s$^{-1}$ at every time instantaneous $t$ and the proton number density $1\,\mathrm{cm}^{-3} < n < 5\,\mathrm{cm}^{-3}$ at ⩾90% of the time duration. There are no strong shear flows or interaction regions in each interval.

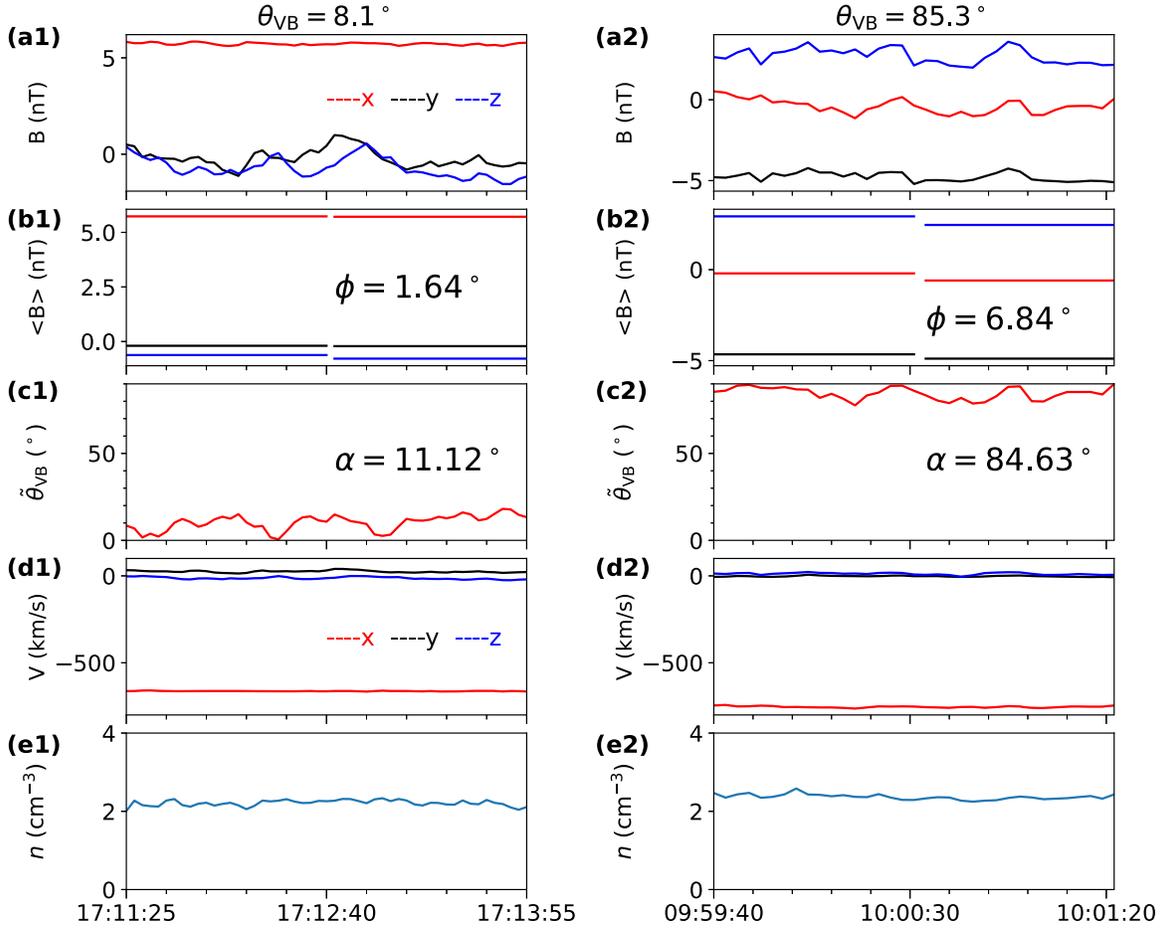

**Figure 2.** Two 150 s subintervals on 2005 October 8. Left: parallel interval with $\theta_{VB} = 8°.1$ and $\phi = 1°.64$. (a1) Time series of magnetic field. Red, black, and blue lines correspond to the $x$, $y$, $z$ components in GSE coordinates. (b1) Two mean magnetic fields of two half intervals. The angle $\phi$ between them is shown. (c1) The instantaneous angle $\tilde{\theta}_{VB}$ between the magnetic field and the velocity. The rms of $\tilde{\theta}_{VB}$ as $\alpha$ is shown. (d1) Time series of velocity. Red, black, and blue lines correspond to the $x$, $y$, $z$ components in GSE coordinates. (e1) Time series of proton number density. Right: perpendicular interval with $\theta_{VB} = 85°.3$ and $\phi = 6°.84$ in the same format as the left panels.

We also require the time stationarity for the subinterval to be averaged as a local background field. We apply the following criterion: for a pair of magnetic field $\boldsymbol{B}_1 = \boldsymbol{B}(t)$, $\boldsymbol{B}_2 = \boldsymbol{B}(t+\tau)$ separated by a time lag $\tau$, we cut the subinterval $\boldsymbol{B}(t')$ ($t' = t, t + \Delta, t + 2\Delta,\ldots,t + \tau$) into two halves with the same time duration of $\tau/2$ and calculate their averages respectively. We define $\phi\,(t,\tau)$ as the angle between the two averages. We only select these pairs with $\phi < 10°$ for further study so that $\boldsymbol{B}(t')$ is time stationary (the stationarity of the average is ensured; Tu & Marsch 1995). We will discuss the criterion of the time stationarity in the next section.

We calculate the local background magnetic field starting at time $t$ and lasting for $\tau$ as

$$\boldsymbol{B}_0(t, \tau) = \langle \boldsymbol{B}(t') \rangle, \quad (1)$$

where $\langle \rangle$ denotes an ensamble time average. The sampling angle $\theta_{VB}$ is the angle between the local background magnetic field $\boldsymbol{B}_0(t, \tau)$ and the solar wind local velocity $\boldsymbol{V}_0(t, \tau)$ (obtained in the same way as $\boldsymbol{B}_0(t, \tau)$). Any angles greater than 90° are reflected below 90°. In the left panel of Figure 2, we show a $\tau = 150$ s subinterval starting at time 17: 11: 25 on 2005 October 8 with $\theta_{VB} = 8°.1$ and $\phi = 1°.64$. Figure 2(b1) shows these two averages $\langle B \rangle$ and the angle $\phi < 10°$. Figure 2(a1) shows the stationary time series of the magnetic field in this subinterval. Time series of velocity and density are shown in Figures 2(d1) and (e1), respectively. We also show a $\tau = 150$ s subinterval starting at time 09: 59: 40 on 2005 October 8 with $\theta_{VB} = 85°.3$ and $\phi = 6°.84$ in the right panel of Figure 2. The time stationarity of both subintervals is pretty well satisfied. The timescale $\tau$ for each subinterval is transferred into the spatial scale as $r(t, \tau) = \tau \times \boldsymbol{V}_0$ under Taylor hypothesis (Taylor 1938; $\boldsymbol{V}_0$ is the average of the velocity in the corresponding subinterval as shown in Figures 2(d1) and (d2)).

We compute the fluctuation

$$\delta \boldsymbol{U}(t, r) = \boldsymbol{U}(t) - \boldsymbol{U}(t+\tau), \quad (2)$$

and obtain the fluctuation power level:

$$\delta U^2(t, r) = \sum_j \delta U_j^2(t, r), \quad (3)$$

where $U$ could be either $B$ or $V$, and $j$ means $x$, $y$, $z$ component in the GSE coordinates. The local parallel and local perpendicular structure functions are calculated as

$$\delta U^2(r, \|) = \frac{1}{n(\theta_{\rm VB})} \sum_{\substack{0° \leqslant \theta_{\rm VB}(t,r) < 10°, \\ \phi(t,r) < 10°}} \delta U^2(t, r), \quad (4)$$

and,

$$\delta U^2(r, \perp) = \frac{1}{n(\theta_{\rm VB})} \sum_{\substack{80° < \theta_{\rm VB}(t,r) \leqslant 90°, \\ \phi(t,r) < 10°}} \delta U^2(t, r), \quad (5)$$

where $n(\theta_{\rm VB})$ denotes the number of the time-stationary subintervals in corresponding bins. We use 66 logarithmically spatial increments to measure the structure function in the range $10^{-4}\,{\rm Mm}^{-1} < k < 10^0\,{\rm Mm}^{-1}$ ($k = 1/r$ is the wavenumber). So that we obtain the parallel ($0° <= \theta_{\rm VB} < 10°$) and perpendicular ($80° < \theta_{\rm VB} <= 90°$) structure functions $\delta U^2(k)$ (of both magnetic field and velocity) in the solar wind turbulence with a time stationary background field ($\phi < 10°$). We then perform least-square regressions to the four structure functions in the range $10^{-2}\,{\rm Mm}^{-1} < k < 10^{-1}\,{\rm Mm}^{-1}$ and get the estimated coefficients and their accompanying standard errors $\sigma$ for 103 fast wind intervals. The quality of the regressions are evaluated by $\sigma$. If $\sigma \geqslant 0.05$, we consider it as a bad regression and remove this interval. 88 intervals with $\sigma < 0.05$ for all four regressions ($\delta B_\|^2$, $\delta B_\perp^2$, $\delta V_\|^2$, $\delta V_\perp^2$) are reserved for the statistical analyses. The coefficients are the scaling indices of the structure functions.

We also analyze $\delta B/B_0$ and $\delta V/V_A$ at $k \approx 10^{-2}\,{\rm Mm}^{-1}$, where $\delta B/B_0$ is the average of $|\delta B|/|\boldsymbol{B}_0|$ and $\delta V/V_A$ is the average of $|\delta V|/(|\boldsymbol{B}_0|/(4\pi m_p n_0)^{1/2})$ within the range $95\,{\rm Mm} < 1/k < 105\,{\rm Mm}$, $n_0$ is the average of proton number density series between two points, shown in Figures 2(e1) and (e2). The results are presented in Section 4.

## 3. Criterion of Time Stationarity

The velocity and magnetic field fluctuations observed by spacecraft in the solar wind include not only the turbulence, but also various structures (Tu & Marsch 1995). Potential structures include current sheets (Li 2007; Miao et al. 2011), the quasi-static convective magnetic field directional turnings (MFDTs; Tu & Marsch 1991, 1992, 1993) and some tangential turnings (TTs) corresponding to re-entrant loops of magnetic field lines in the photosphere (Nakagawa et al. 1989; Nakagawa 1993; Tu et al. 2016), flux tubes that originated at the Sun (Bruno et al. 2001; Borovsky 2008), and interplanetary magnetic flux ropes (Tu et al. 1997). These convective structures may not be considered as turbulence. However, they are more likely to cause higher measured fluctuation amplitudes at larger scales and thus lead to steeper power spectra. Figure 3 is a cartoon showing a possible pair of measurement points that are located separately in two magnetic flux tubes with the subinterval averaged magnetic field direction parallel to the velocity. The two magnetic flux tubes may have originated from different regions on the Sun and the difference between measurements at point 1 and point 2 may not be considered to be local turbulence. Further, the average magnetic field of the time series measured between point 1 and point 2 is parallel to the velocity shown by an arrow. However, this individual measurement shows that the field vector changes its orientation. The field average does not represent the major direction and hence should not be used to determine the sampling angle. To avoid these effects, the intervals influenced by these structures should be rejected in the data analysis.

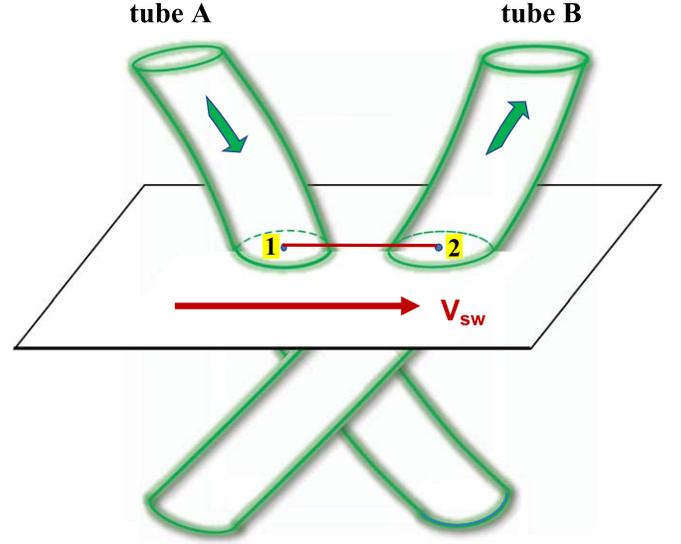

**Figure 3.** Schematic illustration of a rejected subinterval measured from point 1 to point 2. The average field of this subinterval is parallel to the solar wind velocity $V_{\rm SW}$. However, point 1 and point 2 are in different magnetic tubes. The difference of the field between point 1 and point 2 may not represent the turbulence. Therefore, this subinterval should be rejected.

In order to investigate the scaling index anisotropy, one needs to distinguish between the fluctuations of the subintervals with the solar wind velocity parallel and perpendicular, respectively, to the local magnetic field direction, which is calculated as the average of the magnetic field measured in the subinterval (Horbury et al. 2008; Podesta 2009; Luo & Wu 2010; Chen et al. 2012). However, if there is a large structure in the subinterval, the variation is large and the average field direction may not represent the major instantaneous field directions. This effect may mix the parallel and perpendicular measurements. These subintervals should be rejected in the selection process. We calculate $\theta_{\rm VB}$, the angle between the solar wind local velocity and the local background magnetic field averaged in every subinterval. One detects the parallel fluctuations with $\theta_{\rm VB} = 0°$ and the perpendicular fluctuations with $\theta_{\rm VB} = 90°$. For any value with $0° < \theta_{\rm VB} < 90°$, the detected fluctuations are a mix of parallel fluctuations and perpendicular fluctuations. However, a range of $\theta_{\rm VB}$ is allowed to obtain a large sampling for a statistical purpose. Since the perpendicular fluctuation's energy is much larger than the parallel one, we need to limit $\theta_{\rm VB}$ to a small range for the parallel detection measurements, for example, $\theta_{\rm VB} < 10°$. However, if the subintervals contain a large structure or variation, the angles between the instantaneous magnetic field and instantaneous velocity could have a significant amount of values larger than 10°, mixing the parallel fluctuations with the perpendicular fluctuations. The subintervals with large structures or variations should be removed. In the solar wind observation, the larger the scale is, the more contained structures exist and the more nonparallel fluctuations are detected with $\theta_{\rm VB} < 10°$, making the power spectra steeper. To prevent these nonparallel fluctuations from mixing into our measured parallel fluctuations, the time series of magnetic field in these subintervals are required to be time stationary.

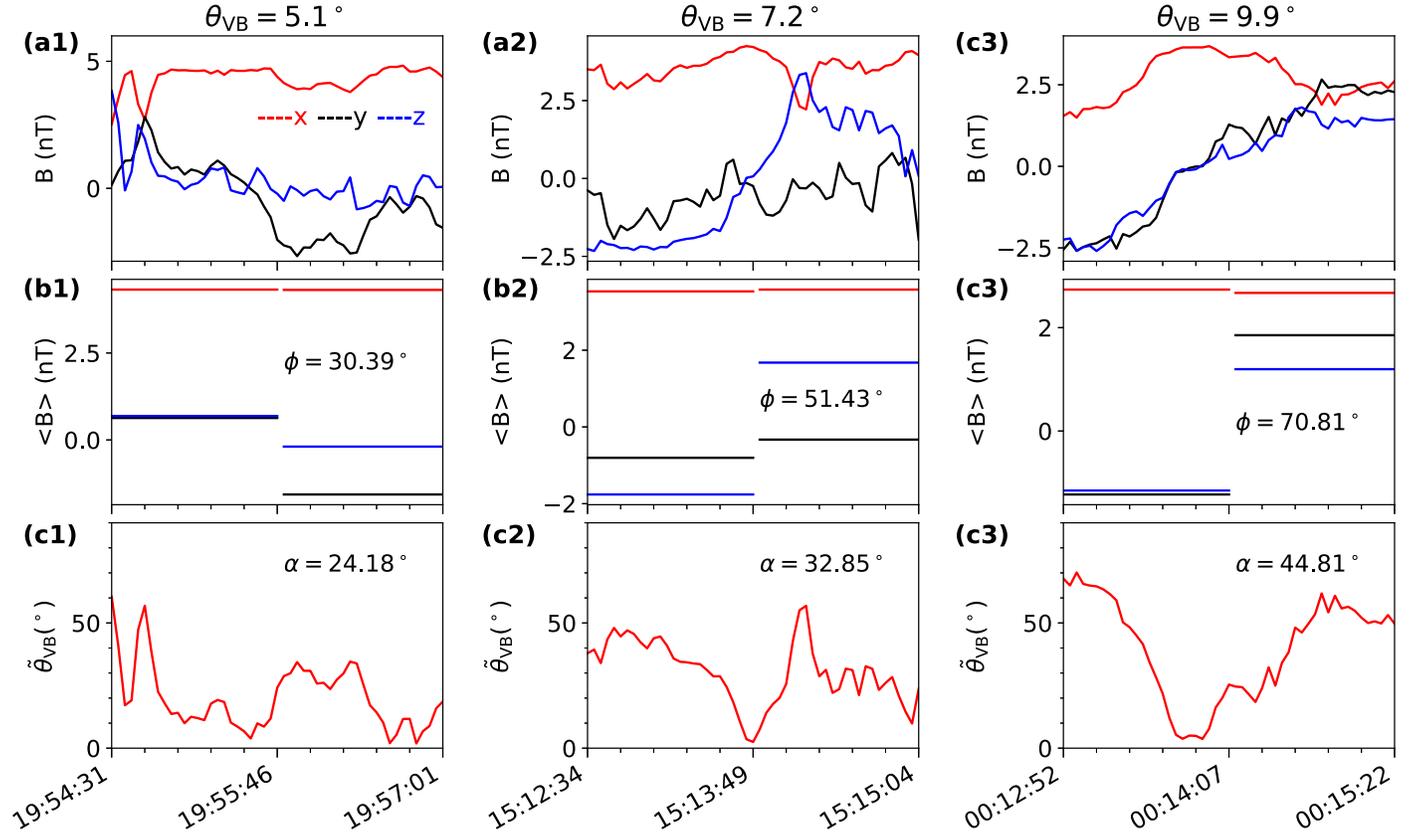

**Figure 4.** Left: a 150 s subinterval on 2005 October 8 with $\theta_{VB} = 5°.1$ (shown in the title) and $\phi = 30°.39$. (a1) Time series of magnetic field. Red, black, and blue lines correspond to the $x$, $y$, $z$ components in GSE coordinates. (b1) Two mean magnetic fields of two half-sub-intervals. The angle $\phi$ between them is shown. (c1) The instantaneous angle $\tilde{\theta}_{VB}$ between the magnetic field and the velocity. The rms of $\tilde{\theta}_{VB}$ as $\alpha$ is shown. Middle: a 150 s subinterval on 2005 October 10 with $\theta_{VB} = 7°.2$ and $\phi = 51°.43$ in the same format as the left panel. Right: a 150 s subinterval on 2005 October 10 with $\theta_{VB} = 9°.9$ and $\phi = 70°.81$ in the same format as the left panel.

We calculate $\phi(t, \tau)$ for each subinterval starting at time $t$ and lasting for $\tau$. $\phi(t, \tau)$ determines the degree of time stationarity. In Figure 4, the first row shows $\boldsymbol{B}(t')$ for three subintervals with $\tau = 150$ s in the left, middle, and right columns, respectively. Figure 4 (a1) shows a current sheet with a reverse in $z$ component lasting for half the time of the subinterval. Figure 4 (a2) shows a current sheet with a reverse in $y$ component lasting for half time of the subinterval. Figure 4 (a3) shows a current sheet with reverses in both $y$ and $z$ components lasting for the whole subinterval. The reverses in the $y$ or $z$ component cause small values of the $y$ or $z$ component of the mean magnetic field and leads to $\theta_{VB} < 10°$, as shown in Figure 3. The second row shows two mean magnetic fields of the two half-sub-intervals as well as the angle $\phi$ between them. The third row shows the instantaneous angle $\tilde{\theta}_{VB}$ between the instantaneous magnetic field and the instantaneous solar wind velocity. From the left to the right, $\phi = 30°.39, 51°.43, 70°.81$. As $\phi$ increases, we can see the instantaneous angle varies more with time and $\alpha$ (the rms of the instantaneous angle) increases from $24°.18$ to $32°.85$, and to $44°.81$, all much larger than $10°$. These suggest the fluctuations of these subintervals with $\theta_{VB} < 10°$ are mixed with the perpendicular fluctuations as a result of the existence of structures, as shown in Figure 3. They should be removed from the data set to calculate the parallel structure function.

We select all the subintervals with $\theta_{VB} < 10°$ and then remove those subintervals with $\phi \geq \phi_c$. We set a series for $\phi_c$ as [90°, 80°, 70°, 60°, 50°, 40°, 30°, 20°, 10°]. We calculate $\alpha$ for every subintervals reserved. We show, in Figure 5(a), the average of $\alpha$ of those subintervals with $\phi \geq \phi_c$ at timescale $\tau = 150$ s. Both the average and the standard deviation of $\alpha$ decrease as $\phi_c$ decreases from $90°$ to $10°$. We also show the average of $\alpha$ of those subintervals with $\phi \in$ the given bin at timescale $\tau = 150$ s in Figure 5(b). There are no subintervals with $80° \leq \phi < 90°$. $\langle\alpha\rangle$ changes from $45.0°$ for these subintervals with $70° \leq \phi < 80°$ to $8.9°$ for these subintervals with $0° \leq \phi < 10°$. The larger $\phi$ is, the larger $\langle\alpha\rangle$ is. We take $\phi < 10°$ as a criterion to select the time stationary subintervals such that the convective structures are rejected and their averaged magnetic field directions represent the major instantaneous field directions in the corresponding intervals. Figure 2 shows that the time stationarity is pretty well satisfied for the $\tau = 150$ s subintervals with $\phi = 1°.64$ and $\phi = 6°.84$.

We accumulate these subintervals with $\theta_{VB} < 10°$ and then remove these subintervals with $\phi \geq \phi_c$. We obtain the magnetic-trace structure functions $\delta B^2_{\phi_c}(k, \parallel)$ for reserved subintervals. We perform least-square regressions to $\delta B^2_{\phi_c}(r, \parallel)$ with a given value of $\phi_c$ in the range $10^{-2}$ Mm$^{-1}$ < $k < 10^{-1}$ Mm$^{-1}$ and get the estimated coefficients (the scaling index). Figure 5(c) shows that the scaling index varies with respect to $\phi_c$ when we remove the subintervals with $\phi \geq \phi_c$. It changes from $-0.81$ with $\phi_c = 90°$ to $-0.64$ with $\phi_c = 10°$. Figure 5(d) shows the magnetic scaling index for $\theta_{VB} < 10°$ with respect to $\phi$ bins when the subintervals with $\phi$ in each bin are accumulated. Beyond $40°$, there are no enough subintervals to obtain a nice power-law structure function. We can see that the scaling index is already close to 0 when $\phi \geq 20°$ and for the bin $10° \leq \phi < 20°$ the scaling is $-0.21$. Figures 5(e) and

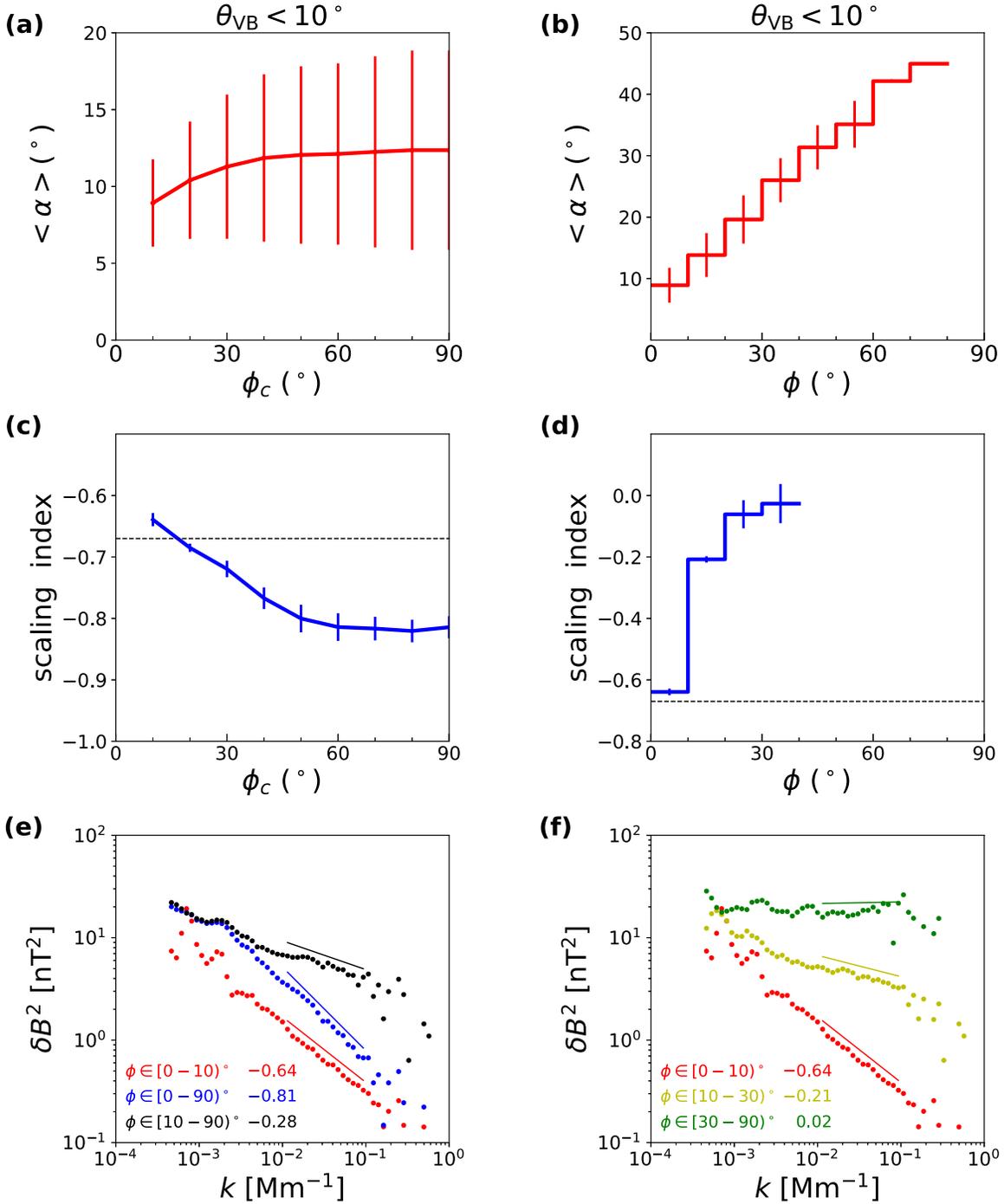

**Figure 5.** Results with different criteria of $\phi$ for those $\theta_{VB} < 10°$ subintervals in the interval (2005 October 8 09:00:00 UT−2005 October 11 16:00:00 UT) shown in Figure 1. (a) Average of $\alpha$ variation at timescale $\tau = 150$ s with respect to $\phi_c$ when the subintervals with $\phi \geqslant \phi_c$ are removed. (b) Average of $\alpha$ variation at timescale $\tau = 150$ s with respect to $\phi$ bins when the subintervals with $\phi$ in each bin are accumulated. (c) Magnetic scaling index with respect to $\phi_c$ when the subintervals with $\phi \geqslant \phi_c$ are removed. (d) Magnetic scaling index with respect to $\phi$ bins when the subintervals with $\phi$ in each bin are accumulated. (e) Magnetic structure functions when the subintervals with $\phi$ in the given bin are accumulated. The bin and the slope of the least-squares regression are shown in the corresponding color. The thin lines are plotted to help visualize the slope. (f) Magnetic structure functions with the same format as panel (e).

(f) demonstrate the magnetic-trace structure functions calculated by the accumulated subintervals under the conditions of $\theta_{VB} < 10°$ and $\phi \in$ the given bin. The red line represents the parallel structure function with $\phi < 10°$, which ensures the time stationarity. It shows a scaling index $-0.64$, close to $-5/3$. Those subintervals with $10 \leqslant \phi < 90°$ are removed and should be removed from the parallel samplings to guarantee the true parallel measurements, since their fluctuations are mixed with the perpendicular fluctuations. The black line shows their structure function with scaling index equal to $-0.28$. The index of 0.02 of the structure functions based on the measurements from the rejected subintervals with $\phi \in [30°, 90°)$ indicates that these fluctuations may be statistically uncorrelated (Bruno et al. 2014). Mixing the subintervals with a stationary background and the subintervals with a nonstationary background, we obtain the blue line. Its scaling index is $-0.81$, steeper than the structure function with a stationary background and is close to the previous studies (Chen et al. 2012).

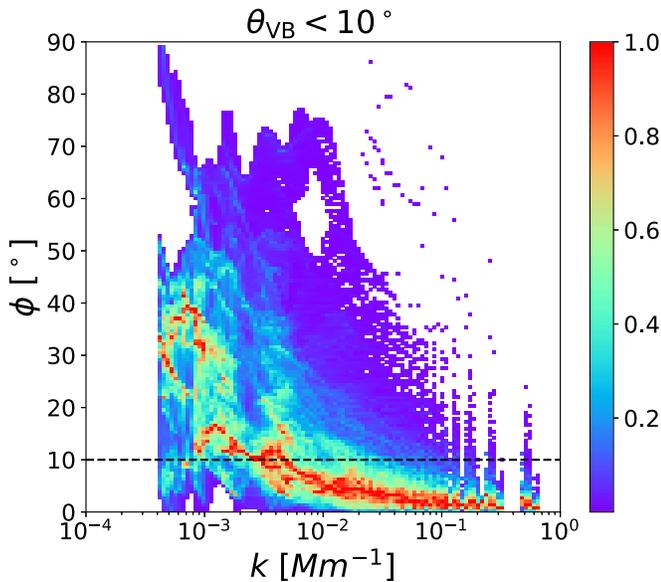

**Figure 6.** Linearly scaled, column-normalized distribution of subintervals in the $\phi - k$ plane for those subintervals with $\theta_{VB} < 10°$ in the interval (2005 October 8 09:00:00 UT–2005 October 11 16:00:00 UT) shown in Figure 1. The dashed line denotes the criterion of $\phi < 10°$.

Figure 6 shows the linearly scaled, column-normalized distribution of subintervals as a function of $\phi$ and $k$. It illustrates that at a large scale, these subintervals with a nonstationary background field account for a significant part of these subintervals with $\theta_{VB} < 10°$ (58% with $\phi < 10°$ for $k \sim 10^{-2}\,\mathrm{Mm}^{-1}$), while at a small scale, these subintervals with a stationary background field play a major role (91% with $\phi < 10°$ for $k \sim 10^{-1}\,\mathrm{Mm}^{-1}$). Therefore, the nonstationary parallel subintervals, suggesting the existence of structures, mainly affect the large-scale detection and lead to a steeper structure function for all the subintervals with $\theta_{VB} < 10°$.

## 4. Results

In the left panel of Figure 7, we present the local parallel and local perpendicular magnetic-trace structure functions for one typical interval (2005 October 8 09:00:00 UT–2005 October 11 16:00:00 UT). The scaling indices are $-0.64$ and $-0.69$ for the local parallel and local perpendicular fluctuations respectively. In the right panel of Figure 7, we show the velocity-trace structure functions. The local parallel and local perpendicular scaling indices are $-0.57$ and $-0.58$. For this interval, we can clearly see the isotropic scaling features of the solar wind turbulence measured locally with stationary background field.

Figure 8 shows the probability density functions of the scaling index calculated for 88 intervals. The average scaling indices are found to be $-0.63 \pm 0.08$ and $0.70 \pm 0.04$ for the local parallel and perpendicular magnetic-trace structure functions, respectively. While for the velocity-trace structure function, the local parallel and perpendicular scaling indices are $-0.47 \pm 0.10$ and $0.51 \pm 0.09$. Note that here the number behind the scaling index is the standard deviation of 88 corresponding scaling indices. It describes the dispersion of the distribution and is different from the standard error $\sigma$ of the estimate coefficients for each fast wind interval. Consider the standard deviation, the probability density functions present isotropic scaling features for both the magnetic field and the velocity. Furthermore, magnetic-trace structure functions are steeper than velocity-trace structure functions.

We show the probability density function of $\delta B/B_0$ and $\delta V/V_A$ at $k \approx 10^{-2}\,\mathrm{Mm}^{-1}$ in Figure 9. $\delta B/B_0$ at $k \approx 10^{-2}\,\mathrm{Mm}^{-1}$ is 0.19 for the parallel fluctuations and 0.29 for perpendicular fluctuations. For $\delta V/V_A$ at $k \approx 10^{-2}\,\mathrm{Mm}^{-1}$, the values are 0.14 and 0.20. The amplitude of the local parallel fluctuations is lower than the amplitude of the perpendicular fluctuations. The magnetic fluctuation amplitude is higher than the velocity fluctuation amplitude of the corresponding parallel and perpendicular conditions.

## 5. Discussion and Conclusions

We show a fast solar wind interval lasting over 3 days without strong shear flows and interaction regions observed by *WIND* spacecraft. The time stationarity of the magnetic field time series in the subintervals is required so that the averaged field direction in the subinterval is representing the major instantaneous field directions and the fluctuation of the magnetic field and the velocity is without influence by convective structures. To guarantee the time stationarity, we reject individual subintervals if the magnetic field in each half differs by more than 10°. This process, then, removes those subintervals with convective structures and prevents the mix of parallel and perpendicular measurements. We determine the local background magnetic field $\boldsymbol{B}_0(t, \tau)$ as the average of the time stationary subintervals starting at time $t$ and lasting for $\tau$. Both the parallel and the perpendicular geometry are guaranteed during the sampling intervals with the stationary background field. The fluctuation power levels with different sampling angles $\theta_{VB}(t, \tau)$ are accumulated and we obtain both parallel and perpendicular second-order structure functions for this whole fast solar wind interval. We find that for magnetic-trace structure functions, the scaling indices are $-0.64 \pm 0.01$ and $0.69 \pm 0.01$ in parallel and perpendicular directions, respectively, and for velocity-trace structure functions they are $-0.57 \pm 0.01$ and $0.58 \pm 0.01$. The scaling indices indicate isotropic scaling features.

We do a statistical analysis with 88 fast solar wind intervals (their time durations are all longer than 2 days). We find that the parallel and perpendicular scaling indices are $-0.63 \pm 0.08$ and $0.70 \pm 0.04$ for the magnetic fluctuations and $-0.47 \pm 0.10$ and $0.51 \pm 0.09$ for the velocity fluctuations. With the selections rejecting those nonstationary subintervals which may be influenced by the convective structures in the solar wind and may mix the parallel and perpendicular measurements, we find the scaling of the structure functions are nearly isotropic, being near $-5/3$ for magnetic field and near $-3/2$ for velocity, which is not consistent with the existing theoretical predictions for the solar wind turbulence. Podesta et al. (2007) found that the magnetic field spectra in the ecliptic plane near 1 au exhibit a scaling index near $-5/3$ and the velocity spectra in the ecliptic plane near 1 au exhibit a scaling index near $-3/2$ using Fourier FFT techniques and Salem et al. (2009) confirmed these scaling laws that the magnetic field scaling index is close to $-5/3$ and the velocity scaling index is close to $-3/2$ after the removal of intermittency. However, they did not investigate the scaling anisotropy. The scaling isotropy we found here is consistent with the isotropic 3D self-correlation level contours of the magnetic field and the velocity in the fast wind shown in Wu et al. (2019a, 2019b).

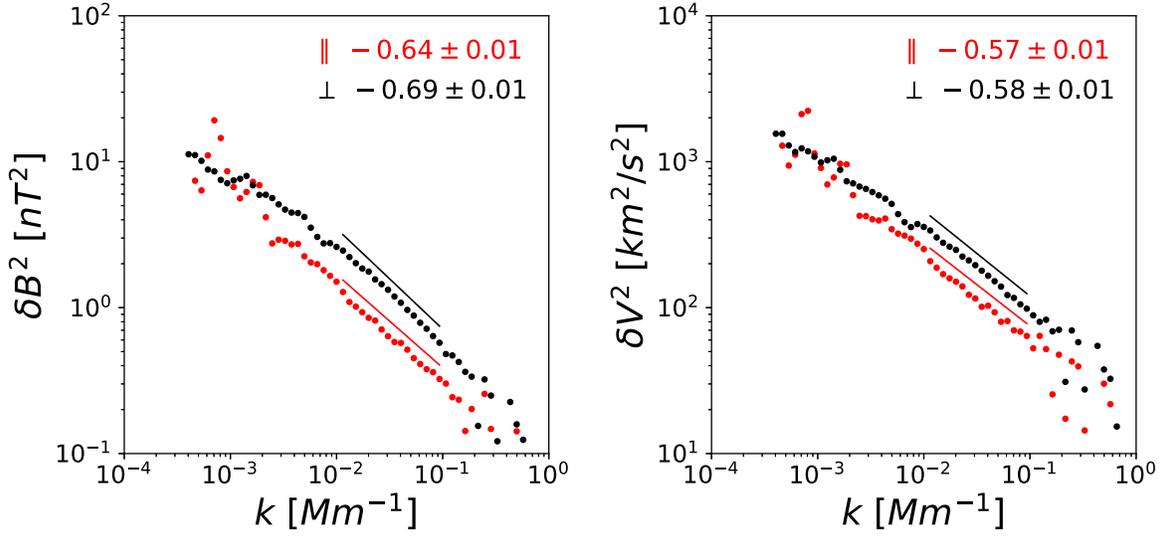

**Figure 7.** Left: magnetic-trace structure functions for a typical fast wind interval (2005 October 8 09:00:00 UT–2005 October 11 16:00:00 UT). Red and black colors correspond to the local parallel and local perpendicular conditions. The scaling indices (the estimated slope of a least-squares regression to the corresponding structure functions in the range $10^{-2}\,\mathrm{Mm}^{-1} < k < 10^{-1}\,\mathrm{Mm}^{-1}$) and their standard errors are shown in the corresponding colors. The thin lines are plotted to help visualize the slope. Right: velocity-trace structure functions for the same interval with the same format as the left panel.

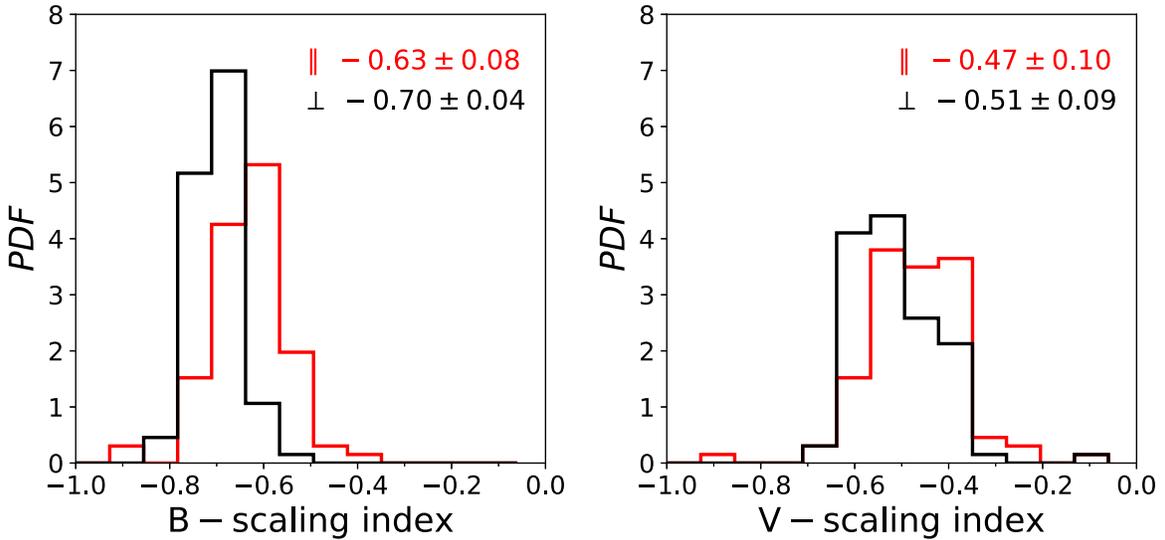

**Figure 8.** Left: probability density function of the magnetic-trace structure function scaling index. Red and black colors correspond to the local parallel and local perpendicular conditions. The averages and standard deviations of the scaling indices are shown in corresponding colors. Right: probability density function of the velocity-trace structure function scaling index with the same format as the left panel.

Wang et al. (2015) selected the low-amplitude time series so that the time series are time stationary and investigated their scaling anisotropy. They found that the parallel magnetic scaling index is $-1.67$ and the perpendicular index is $-1.46$. Wang et al. (2016) investigate the scaling anisotropy based on the wavelet analysis under the requirement that the time stationarity is better ensured and found that the parallel magnetic scaling index is $-1.75$ and the perpendicular index is $-1.65$. However, compared with the values 0.05 for the parallel fluctuations at a timescale of 6 minutes in Wang et al. (2015) and 0.04 for the parallel fluctuations at a timescale of 100 s in Wang et al. (2016), we can conclude that we obtain the scaling features for the moderate-amplitude fluctuations that the $\delta B/B_0$ at $k \approx 10^{-2}\,\mathrm{Mm}^{-1}$ is 0.19 for the parallel fluctuations and 0.29 for perpendicular fluctuations. Both Wang et al. (2015) and Wang et al. (2016) did not study the scaling anisotropy for the velocity. Wang et al. (2014) found the scaling isotropy, $-1.63 \pm 0.02$ for magnetic field and $-1.56 \pm 0.02$ for velocity, after the removal of the data points with large wavelet coefficients (they defined such data as intermittency). Their results are consistent with our results. However, they did not show the detailed reason why the data points with large wavelet coefficients should be removed. Our work using the structure function analysis is more precise, and we analyze whether parallel and perpendicular fluctuations can be distinguished from each other for every subinterval and reject all those that cannot. We also reserve a larger part of the data and the fluctuation power is higher than that in Wang et al. (2014). After this paper was submitted, we found that Telloni et al. (2019) was published showing that the parallel magnetic spectral index is characterized by $-5/3$ after eliminating the possible role of nonstationarity and large-scale structures using Hilbert spectral analysis. They did not show the isotropic scaling index for the velocity.

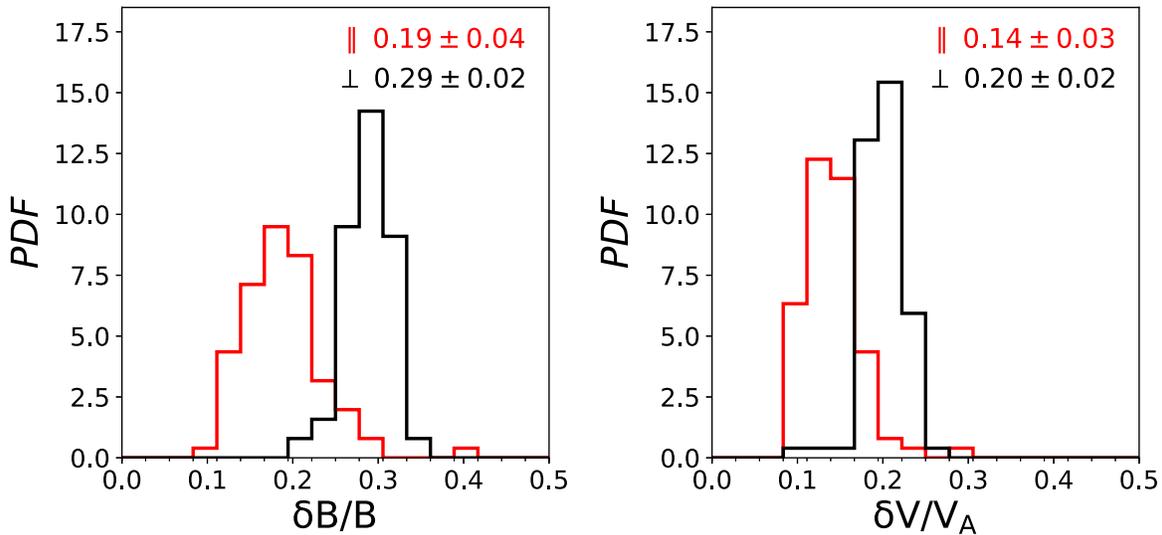

**Figure 9.** Left: probability density function of $\delta B/B_0$ at $k = 10^{-2}$ Mm$^{-1}$. Red and black colors correspond to the local parallel and local perpendicular conditions. The averages and standard deviations of the scaling indices are shown in the corresponding colors. Right: probability density function of $\delta V/V_A$ at $k \approx 10^{-2}$ Mm$^{-1}$ with the same format as the left panel.

We thank the CDAWEB for access to the *WIND* data and Dr. Daniel Verscharen for helpful discussion. This work at Peking University and Beihang University is supported by the National Natural Science Foundation of China under contract Nos. 41674171, 41874199, 41574168, 41874200, 41774183, and 41861134033.

## Appendix

The *WIND* 3DP plasma instrument suffers from digitization at the smallest scales due to compression in its data, especially in the *x* component. Here we use the sum of the *y* and *z* components of the fluctuation's power level instead of the trace for velocity. We show the sum of the *y* and *z* components of the fluctuations power level as a function of scale in Figure 10 for the same interval as in Figure 7. The scaling (0.59, 0.60) is very close to the scaling (0.57, 0.58) of the velocity-trace structure functions. We do the same statistical analysis and find that the average of the scaling is $0.49 \pm 0.11$ measured in the parallel direction and $0.51 \pm 0.09$ measured in the perpendicular direction, suggesting that the digitization of the *x* component effect is negligible to analyze the scaling in the region $10^{-2}$ Mm$^{-1} < k < 10^{-1}$ Mm$^{-1}$. This is reasonable because the digitization is mainly at the smallest scales and we perform the regression above $k = 10^{-1}$ Mm$^{-1}$ which corresponds to larger than 12 s. The digitization may also affect the transformation of the time lag to the spatial lag. However, this effect will not cause a difference of the background velocity larger than 2% and thus is also neglectable.

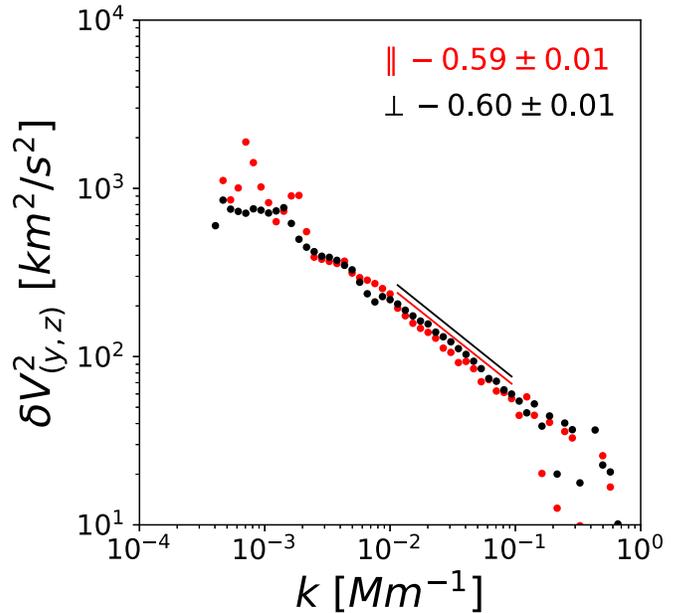

**Figure 10.** Sum of the *y* and *z* components of the velocity fluctuation's power level for the same typical interval and with the same format as in Figure 7.

## ORCID iDs

Honghong Wu https://orcid.org/0000-0003-0424-9228
Chuanyi Tu https://orcid.org/0000-0002-9571-6911
Jiansen He https://orcid.org/0000-0001-8179-417X
Liping Yang https://orcid.org/0000-0003-4716-2958
Linghua Wang https://orcid.org/0000-0001-7309-4325